\documentclass[12pt]{iopart}

\usepackage{graphicx}
\usepackage{amssymb}
\begin{document}
\title{Coupled harmonic oscillators and their quantum entanglement}
\author{D.\,N. Makarov}

\address{Federal Center for Integrated Arctic Research, Russian Academy of Sciences, nab. Severnoi Dviny 23, 163000, Arkhangelsk, Russia.}

\address{Northern (Arctic) Federal University, nab. Severnoi Dviny 17, 163002, Arkhangelsk, Russia.}
\ead{makarovd0608@yandex.ru}
\vspace{10pt}
\begin{indented}
\item[]October 2017
\end{indented}
\begin{abstract}
A system of two coupled quantum harmonic oscillators with the Hamiltonian ${\hat H}=\frac{1}{2}\left(\frac{1}{m_1}{\hat p}^{2}_1 + \frac{1}{m_2}{\hat p}^{2}_2+A x^2_1+B x^2_2+ C x_1 x_2\right)$ can be found in many applications of quantum and nonlinear physics, molecular chemistry, and biophysics. The stationary wave function of such a system is known, but its use for the analysis of quantum entanglement is complicated because of the complexity of computing the Schmidt modes. Moreover, there is no exact analytical solution to the nonstationary Schrodinger equation ${\hat H}\Psi=i\hbar\frac{\partial \Psi}{\partial t}$ and Schmidt modes for such a dynamic system. In this paper we find a solution to the nonstationary Schrodinger equation; we also find in an analytical form a solution to the Schmidt mode for both stationary and dynamic problems. On the basis of the Schmidt modes, the quantum entanglement of the system under consideration is analyzed. It is shown that for certain parameters of the system, quantum entanglement can be very large.
\end{abstract}
\section{Introduction}
The behavior of systems that contain coupled harmonic oscillators is currently an area of very active research; this interest is primarily due to the fact that models of such systems are encountered in many applications of quantum and nonlinear physics \cite{Fano_1957}-\cite{Paz_2008}, molecular chemistry \cite{Ikeda_1999}-\cite{Delor_2017} and biophysics \cite{Romero_2014}-\cite{Halpin_2014}. In quantum physics specifically, this interest is because of the quantum entanglement for such a system. In particular, using entangled states, it is possible to explain quantum communication protocols, such as quantum cryptography \cite{Ekert_1991}, quantum-coding \cite{Bennett_1992}, quantum computing algorithms \cite{Shor_1995} and quantum state teleportation \cite{Aspect_1981,Samuel_1998}. On the other hand, physical models of coupled harmonic oscillators have been used in many studies in physics, such as the Lee model in quantum field theory \cite{Schweber_1961}-\cite{Han_1990}. There are also models in which one of the oscillator variables is not observed \cite{Fano_1957}. These physical models are examples of the remainder of the Feynman universe \cite{Han_Kim_1999,Feynman_1972}. A Hamiltonian with coupled oscillators is used to study vibronic molecules (for example, \cite{Ikeda_1999}) in molecular chemistry, and a similar Hamiltonian is used in biophysics, to explain the process of photosynthesis \cite{Romero_2014}-\cite{Halpin_2014}. 

Of particular interest is the system with the Hamiltonian
\begin{equation}
{\hat H}=\frac{1}{2}\left(\frac{1}{m_1}{\hat p}^{2}_1 + \frac{1}{m_2}{\hat p}^{2}_2+A x^2_1+B x^2_2+ C x_1 x_2\right),
\label{1}
\end{equation}
where ${\hat p}_k=-i\hbar \frac{\partial}{\partial x_k}$ (where $k=(1,2)$) is the momentum operator.
In spite of the fact that the solution of the stationary Schrodinger equation with the Hamiltonian (\ref{1}) is known \cite{Han_Kim_1999}, there are a number of problems, one being the complexity of calculating and analyzing quantum entanglement for such a system. Usually, quantum entanglement is analyzed for the ground state of the oscillators \cite{Han_Kim_1999} because the Schmidt modes $ \lambda_k $ \cite{Ekert_1995}, which are used in the analysis of quantum entanglement, are not calculated in general form. In connection with the importance of the system with the Hamiltonian (\ref{1}), this problem is topical and requires a solution. Also, the parameterization \cite{Han_Kim_1999}, which is usually used in solving the Schrodinger equation (\ref{1}), is difficult to analyze and needs to be simplified. In addition to these problems, there is also a need for an analytical solution to the nonstationary Schrodinger equation with the Hamiltonian (\ref{1}), as well as an analysis of the solution obtained for quantum entanglement using the Schmidt modes. All the above problems are considered in this work and solved. It is also shown that for certain parameters of the system there can be a large quantum entanglement. All results are presented in an analytical form.

\section{Quantum entanglement for a stationary system}
Consider a system with Hamiltonian (\ref{1}). In order to solve the stationary Schrodinger equation with the Hamiltonian (\ref{1}), it must first be brought to diagonal form. Details of the diagonalization of the Hamiltonian (\ref{1}) can be found, for example, in \cite{Han_Kim_1999}, the result being
\begin{equation}
{\hat H}=\frac{1}{2 M}\left({\hat p}^{2}_1 + {\hat p}^{2}_2+\frac{K}{2}\left(e^{\eta}y^2_1 + e^{-\eta}y^2_2  \right) \right),
\label{2}
\end{equation}
where $y_1=x_1\cos \alpha-x_2 \sin \alpha$, $y_2=x_1\sin \alpha + x_2 \cos \alpha$, ${\hat p}_k=-i\hbar \frac{\partial}{\partial y_k}$, $M=\sqrt{m_1 m_2}$, and the parameters $K$, $\alpha$ and $\eta$ take the form
\begin{equation}
K=\sqrt{A B-\frac{C^2}{4}}, e^{\eta}= \frac{A + B + \frac{A-B}{|A-B|}\sqrt{\left( A-B \right)^2+C^2 }}{\sqrt{4 A B -C^2}}, \tan \left( 2\alpha \right) =\frac{C}{B-A} .
\label{3}
\end{equation}
It should be added that in \cite{Han_Kim_1999} and in many other studies $e^{\eta} $ is used in expression (\ref{3}) without the parameter $\frac{A-B}{|A-B|}$. This, of course, is not mathematically correct, so this refinement should be taken into account (although it can be ignored for $ A>B $). There is also a singular point for $A = B$, where $e^{\eta} $ is not de ned and can take two values.

Usually, just such a parameterization (\ref{3}) is used in many problems of the system under consideration. In fact, this is not a simple parameterization, and therefore it is not the most convenient for analyzing the results. Using the properties of trigonometric functions one can obtain the Hamiltonian (\ref{2}) in the form
\begin{equation}
{\hat H}=\frac{1}{2M}\left({\hat p}^{2}_1 + {\hat p}^{2}_2 \right) + \frac{1}{2}\left( A^{'} y^2_1 + B^{'} y^2_2\right)  ,
\label{5}
\end{equation}
where
\begin{equation}
A^{'}=A-\frac{C}{2} \tan \alpha ,~~ B^{'}=B+\frac{C}{2} \tan \alpha . 
\label{6}
\end{equation}
In the expression (\ref{6}) $\tan\alpha \in (-1,1) $, since $\tan \alpha =\frac{\epsilon}{|\epsilon|}\sqrt{\epsilon^2+1}-\epsilon$, where $\epsilon=\frac{B-A}{C}$, and hence $\alpha \in (-\pi/4, \pi/4) $. Thus, the angle $\alpha$ is bounded from $(-\pi/4, \pi/4)$, which must be taken into account for the analysis of the system under consideration. Further, for convenience, we pass to the system of units, where $M=1,\hbar=1$ (similar to the atomic system of units). Let us write out the eigenvalue of the energy of the system and its wave function, we obtain
\begin{equation}
E_{n,m}= \sqrt{A^{'}}\left( n+\frac{1}{2}\right)+ \sqrt{B^{'}}\left(m +\frac{1}{2}\right) , 
\label{7}
\end{equation}
\begin{equation}
\Psi_{n,m}(y_1,y_2)= c_n c_m e^{- \frac{\sqrt{A^{'}}}{2}y^2_1}e^{- \frac{\sqrt{B^{'}}}{2}y^2_2} H_n\left( {A^{'}}^{1/4} y_1 \right)H_m\left({B^{'}}^{1/4} y_2 \right)  , 
\label{8}
\end{equation}
where $c_n={A^{'}}^{1/8}/\sqrt{\sqrt{\pi}n!2^n } $, $c_m={B^{'}}^{1/8}/\sqrt{\sqrt{\pi}m!2^m } $, and $n,m$ are quantum numbers.

From Schmidt's theorem \cite{Ekert_1995,Grobe_1994}, it is known that the wave function of pure states connected to each other by systems 1 and 2 can be expanded in the form $\Psi = \sum_{k}\sqrt{\lambda_k}u_{k}(x_1)v_{k}(x_2) $ where: $u_{k}(x_1)$ is the wave function of the pure state 1 of the system; $v_{k}(x_2)$ is the wave function of the pure state 2 of the system; and $\lambda_k$ is the Schmidt mode, which is the eigenvalue of the reduced density matrix, i.e. $\rho_1(x_1,x^{'}_1)=\sum_{k}\lambda_k u_{k}(x_1)u^{*}_{k}(x^{'}_{1}) $ or $\rho_2(x_2,x^{'}_{2})=\sum_{k}\lambda_k v_{k}(x_2)v^{*}_{k}(x^{'}_{2}) $. If we find the Schmidt mode $\lambda_k$, we can then calculate the measure of the quantum entanglement of the system. To do this, various measures of entanglement can be used, for example, the Schmidt parameter \cite{Ekert_1995,Grobe_1994} $K=\left(\sum_{k}\lambda^2_k \right)^{-1} $ or Von Neumann entropy \cite{Bennett_1996,Casini_2009} $S_N=-\sum_{k} \lambda_k \ln \left(\lambda_k \right) $. The main difficulty in calculating quantum entanglement is the search for $\lambda_k$ of the system under consideration, and this parameter is considered below.

We decompose $\Psi_{n,m}(y_1,y_2)=\sum_{k,p}A^{k,p}_{n,m}\phi_{k}(x_1)\varphi_{p}(x_2)$, where $\phi_{k}(x_1)$ and $\varphi_{p}(x_2)$  are the wave functions of 1 and 2 of the unbound system, respectively, and $A^{k,p}_{n,m}$ are the coefficients of expansion. Using the orthogonality condition, these coefficients can be found using $A^{k,p}_{n,m}=\left\langle \Psi_{n,m}(y_1,y_2)|\phi_{k}(x_1)\varphi_{p}(x_2)\right\rangle $. 
Next, a simplification can be made, connected with the fact that the parameter $C$ in expression (\ref{8}) will be a small value, i.e. $C<<A, C<<B$. Indeed, in realistic models, the relationship between systems is always many times less than the connections within these systems. It can be seen from expressions (\ref{6}) and (\ref{8}), that for $C<<A, C<<B$ the quantum entanglement of the system will be significant in the case when $\epsilon$ is of finite size. Since  $C<<1$, and $\epsilon$  is a finite (not small) quantity, we obtain $A \approx B$, and $ A-B$ is less than or of order $C$.
We write out expression (\ref{8}), retaining the basic terms under the condition $C<<1$ and $ A-B \lesssim C $
\begin{eqnarray}
\Psi_{n,m}(y_1,y_2)= \frac{{A}^{1/4}}{\sqrt{\pi n!2^n m!2^m}}e^{- \frac{\sqrt{A}}{2}\left(x_1\cos \alpha - x_2 \sin \alpha \right)^2}e^{- \frac{\sqrt{A}}{2}\left(x_1\sin \alpha + x_2 \cos \alpha\right)^2}\times
\nonumber\\ 
\times H_n\left( {A}^{1/4} \left(x_1\cos \alpha - x_2 \sin \alpha \right)  \right)H_m\left({A}^{1/4} \left(x_1\sin \alpha + x_2 \cos \alpha \right)  \right). 
\label{9}
\end{eqnarray}
It can be seen from expression(\ref{9}) that quantum entanglement can be large, since $\alpha \in (-\pi/4, \pi/4)$, in spite of the fact that $C<<A, C<<B$. 

In order to calculate $A^{k,p}_{n,m}$ with the wave function (\ref{9}), the results of \cite{Makarov_2017_adf} can be used, where an integral of this kind was calculated. As a result, we obtain
\begin{eqnarray}
A^{k,p}_{n,m}=\frac{\mu^{k+n}\sqrt{m!n!}}{(1+\mu^2)^{\frac{n+m}{2}}\sqrt{k!p!}}P^{(-(1+m+n), m-k)}_{n}\left(-\frac{2+\mu^2}{\mu^2} \right) ,
\label{10}
\end{eqnarray}
where $P^{(b,c)}_{a}(x)$ is the Jacobi polynomial, and $\mu=\tan \alpha$. In the expression (\ref{10}), as was shown in \cite{Makarov_2017_adf} in calculating the integral, the condition $k+p=m+n$ is fulfilled.
Further, after finding the reduced density matrices, we obtain
\begin{eqnarray}
\rho_1(x_1,x^{'}_1)=\sum^{m+n}_{k=0}|A^{k,m+n-k}_{n,m}|^2 \phi_{k}(x_1)\phi^{*}_{k}(x^{'}_1),
\nonumber\\ 
\rho_2(x_2,x^{'}_2)=\sum^{m+n}_{k=0}|A^{k,m+n-k}_{n,m}|^2 \varphi_{m+n-k}(x_2)\varphi^{*}_{m+n-k}(x^{'}_2) .
\label{11}
\end{eqnarray}
As a result, we obtain the Schmidt mode in the form
\begin{eqnarray}
\lambda_k = \frac{\mu^{2(k+n)}m!n!}{\left( 1+\mu^2\right)^{m+n}k!(m+n-k)! } \left( P^{-(1+m+n,m-k)}_{n}\left(-\frac{2+\mu^2}{\mu^2} \right) \right)^2 . 
\label{12}
\end{eqnarray}
It should be noted that expression (\ref{12}) is not a discontinuous function, since it depends on $\mu^2$, in spite of the fact that $\mu$, at $\epsilon = 0$ is a discontinuous function. As a result, for example, the Von Neumann entropy and the Schmidt parameter are, respectively, $S=-\sum^{m+n}_{k=0}\lambda_k \ln \lambda_k$, $K=\left(\sum^{m+n}_{k=0}\lambda^2_k \right)^{-1} $. 
As an example, in figure 1 we present the graphs for $S$ and $K$ depending on the parameter $\mu$ for various combinations of quantum numbers $(m,n)$. Since the Schmidt mode $\lambda_k$  is an even function of $\mu$, it is sufficient to change the parameter $\mu$  within $\mu \in (0,1)$. 
\begin{figure}[]
\begin{minipage}[h]{0.49\linewidth}
\center{\includegraphics[angle=0,width=1\textwidth, keepaspectratio]{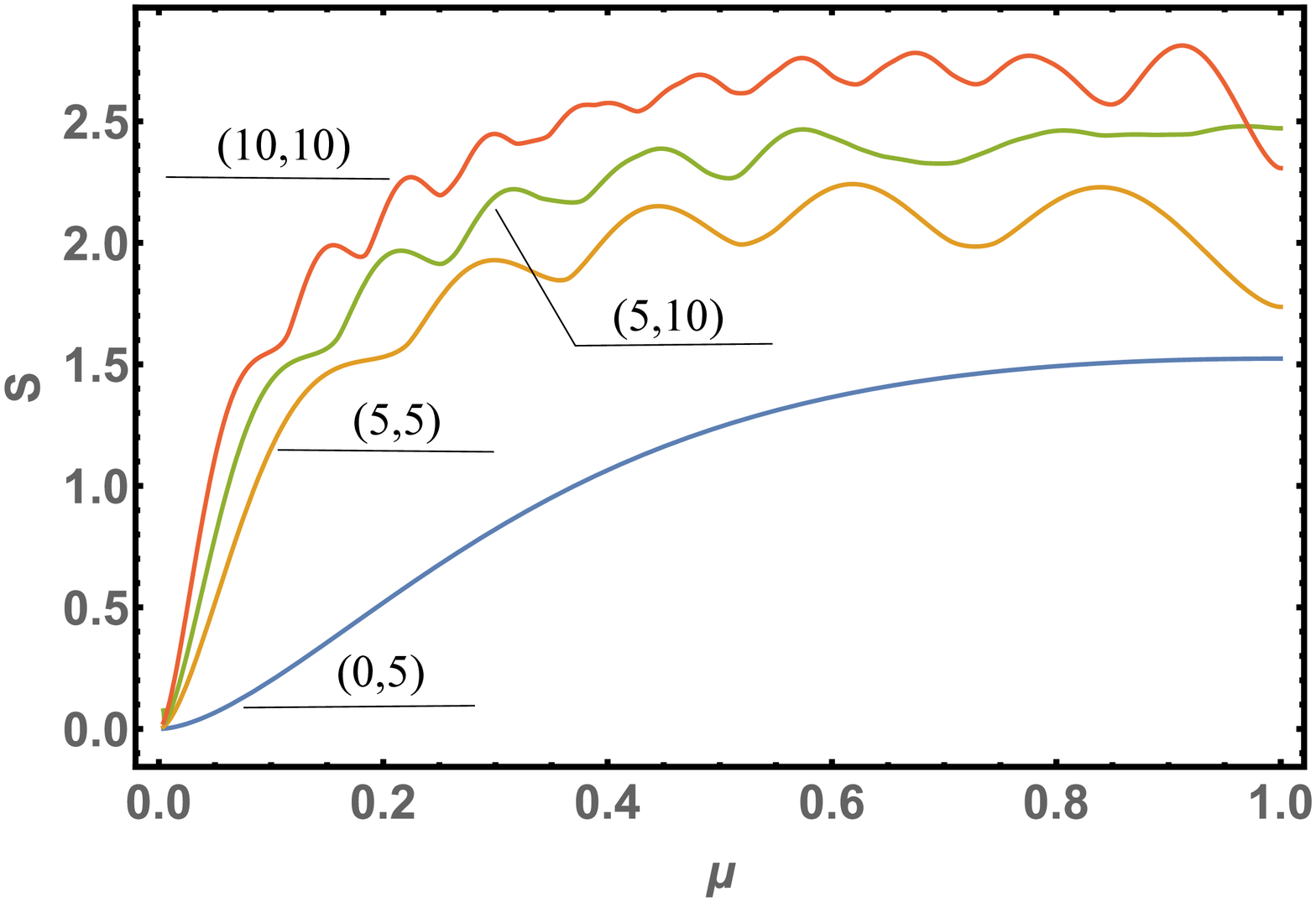}} 
\end{minipage}
\hfill
\begin{minipage}[h]{0.49\linewidth}
\center{\includegraphics[angle=0,width=1\textwidth, keepaspectratio]{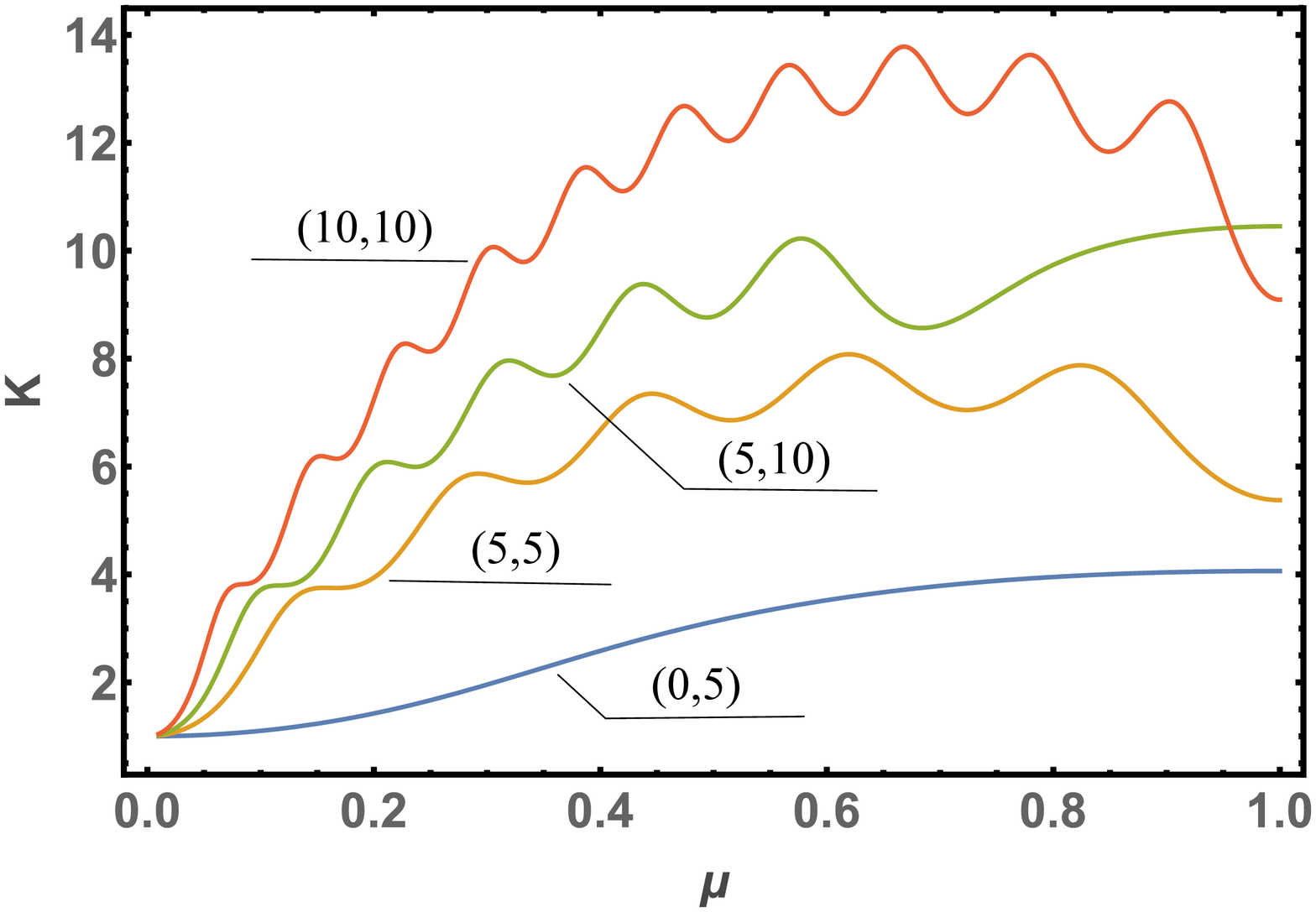}} 
\end{minipage}
\center{\caption[fig1]{Results of calculations of quantum entanglement for the Von Neumann entropy $ S $ and the Schmidt parameter $ K $.}}
\label{fig1}
\end{figure}

We should also add that for $n=0$, using the known expression for the sum \cite{Prudnikov_T3}, the Schmidt parameter is obtained in the form
\begin{eqnarray}
K=\frac{\left( 1+\mu^2 \right)^{2 m} }{{_2}F_1\left(-m,-m;1,\mu^4 \right) },  
\label{13}
\end{eqnarray}
where ${_2}F_1\left(a,b;c,x \right)$ is a hypergeometric function. Expression (\ref{13}) for $ K $ is an increasing function of $ m $ for finite $ \mu $, so the quantum entanglement of such a system is unlimited. In general, quantum entanglement can reach very large values at large quantum numbers $ m, n $.

\section{Quantum entanglement of the dynamic system}

Consider the non-stationary Schrodinger equation ${\hat H}\Psi=i\hbar\frac{\partial \Psi}{\partial t}$, with the Hamiltonian (\ref{1}). We assume that the system at the initial instant of time $t = 0$ was in the states $ |s_1\rangle, |s_2\rangle $, which are the eigenfunctions of unconnected oscillators whose quantum numbers are $s_1,s_2 $. The Hamiltonian for such an unbound system will be in the form
\begin{equation}
{\hat H}_{t=0}=\frac{1}{2}\left(\frac{1}{m_1}{\hat p}^{2}_1 + \frac{1}{m_2}{\hat p}^{2}_2+A x^2_1+B x^2_2\right).
\label{14}
\end{equation}
Making the substitutions of the variable, analogous to the stationary Schrodinger equation, and reapplying the same system of units $M=1,\hbar =1$, we obtain
\begin{equation}
\Psi(x_1,x_2,t)=\sum_{n,m}a_{n,m}e^{-i{E_{n,m}}t}\Psi_{n,m}(y_1,y_2),
\label{15}
\end{equation}
where $E_{n,m}$ is determined by expression (\ref{7}), and $\Psi_{n,m}(y_1,y_2)$ by expression (\ref{8}) for $y_1=x_1\cos \alpha-x_2 \sin \alpha$, $y_2=x_1\sin \alpha + x_2 \cos \alpha$.
As with the previous problem, if we assume that the parameter $C$ within expression (\ref{8}) is a small value, i.e. $C<<A, C<<B$, then $\Psi_{n,m}(y_1,y_2)$ for $C<<A, C<<B$  is defined by expression (\ref{9}). The coefficient $a_{n, m}$ can be easily found by knowing the wave function of the system at $t = 0$. As a result, we obtain an integral analogous to that considered earlier, where $a_{n, m} = A^{s_1,s_2}_{n,m} $ and is defined by expression (\ref{10}). It should be added that, as with the case considered above, the condition $s_1+s_2=n+m$ will be satisfied.

In order to find the Schmidt modes of the problem under consideration, we have to expand $\Psi (x_1,x_2,t)$ in the form
\begin{equation}
\Psi(x_1,x_2,t)=\sum_{k,p}c_{k,p}(t)\phi_{k}(x_1,t)\varphi_{p}(x_2,t),
\label{16}
\end{equation}
where $\phi_{k}(x_1,t) $ and $\varphi_{p}(x_2,t)$ are the eigenfrequencies of the unrelated systems 1 and 2, respectively, and $c_{k,p}(t) $ are the expansion coefficients. It is not difficult to find the coefficient $c_{k, p}(t) $ using the orthogonality condition and the integral already considered
\begin{equation}
c_{k,p}(t)=\sum^{s_1+s_2}_{n=0}A^{s_1,s_2}_{n,s_1+s_2-n}A^{*{k,p}}_{n,s_1+s_2-n}e^{-i{\Delta E_{n,s_{1} + s_{2}-n}}t}  ,
\label{17}
\end{equation}
where
\begin{equation}
\Delta E_{n,s_{1} + s_{2}-n} =E_{n,s_{1} + s_{2}-n}-\sqrt{A}\left(s_1+\frac{1}{2}\right) -\sqrt{B}\left( s_2 +\frac{1}{2}\right)  .
\label{18}
\end{equation}
In expression (\ref{17}), proceeding from the properties of the integral considered in \cite{Makarov_2017_adf}, it should be noted that $ k + p = s_1 + s_2 $. Then, using the density matrix and the Schmidt mode definition, as the latter is an eigenvalue of the reduced density matrix, it is easy to obtain  $ \lambda_k(t)=\left| c_{k,s_1 + s_2 -k}(t)\right|^2 $. 
\begin{figure}[h]
\begin{minipage}[h]{0.49\linewidth}
\center{\includegraphics[angle=0,width=1\textwidth, keepaspectratio]{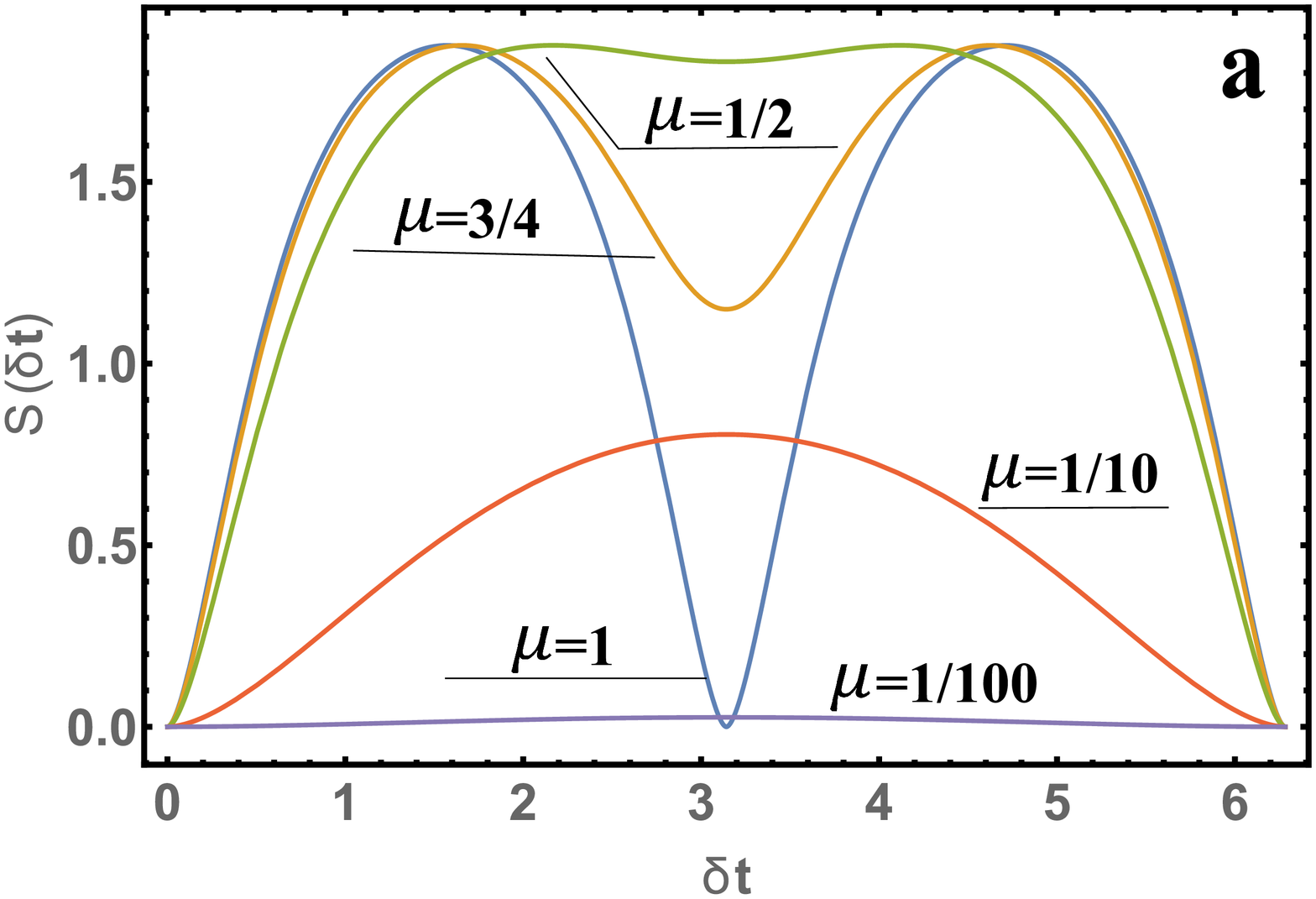}} \\
\end{minipage}
\hfill
\begin{minipage}[h]{0.49\linewidth}
\center{\includegraphics[angle=0,width=1\textwidth, keepaspectratio]{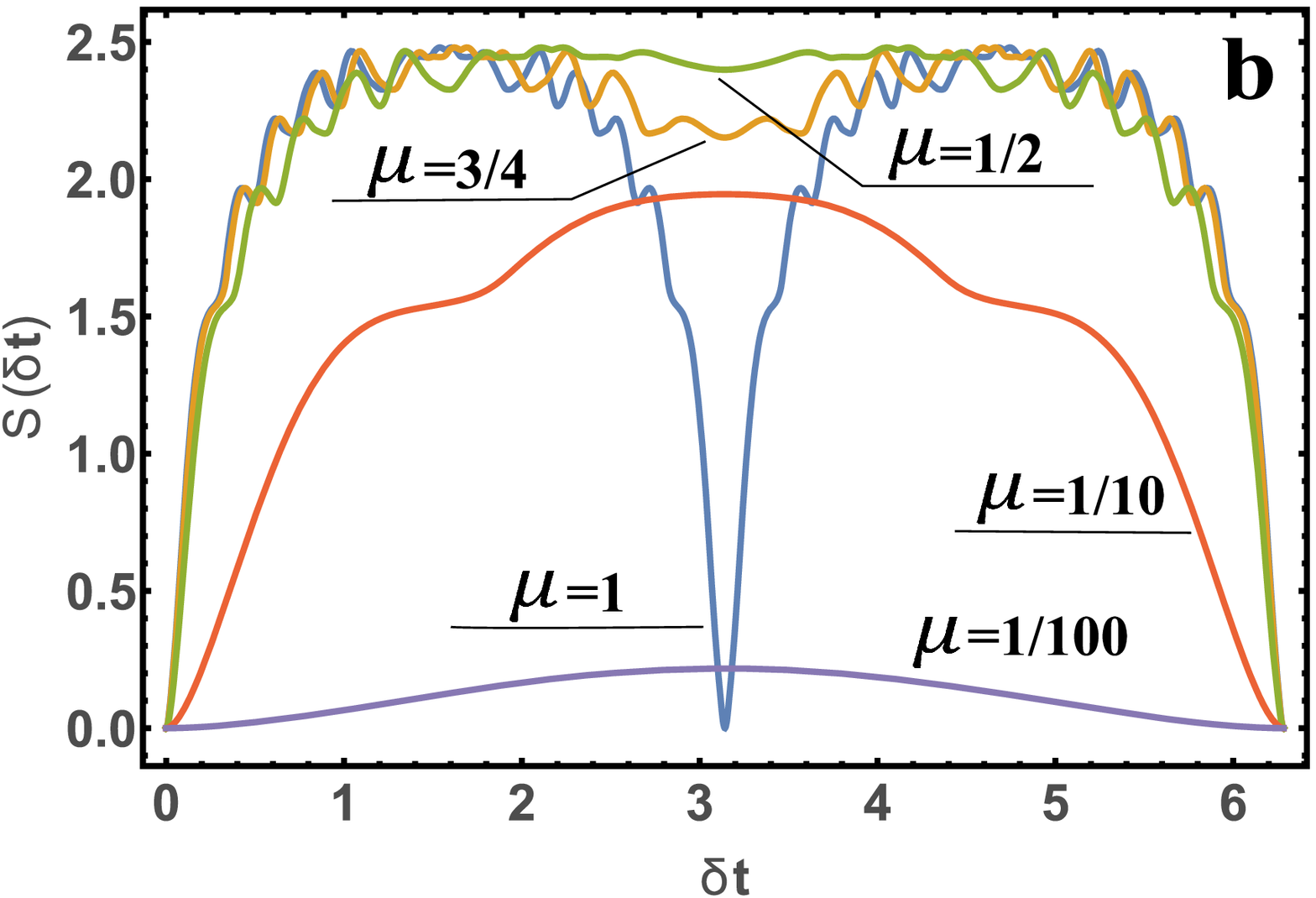}} \\ 
\end{minipage}
\hfill
\begin{minipage}[h]{0.49\linewidth}
\center{\includegraphics[angle=0,width=1\textwidth, keepaspectratio]{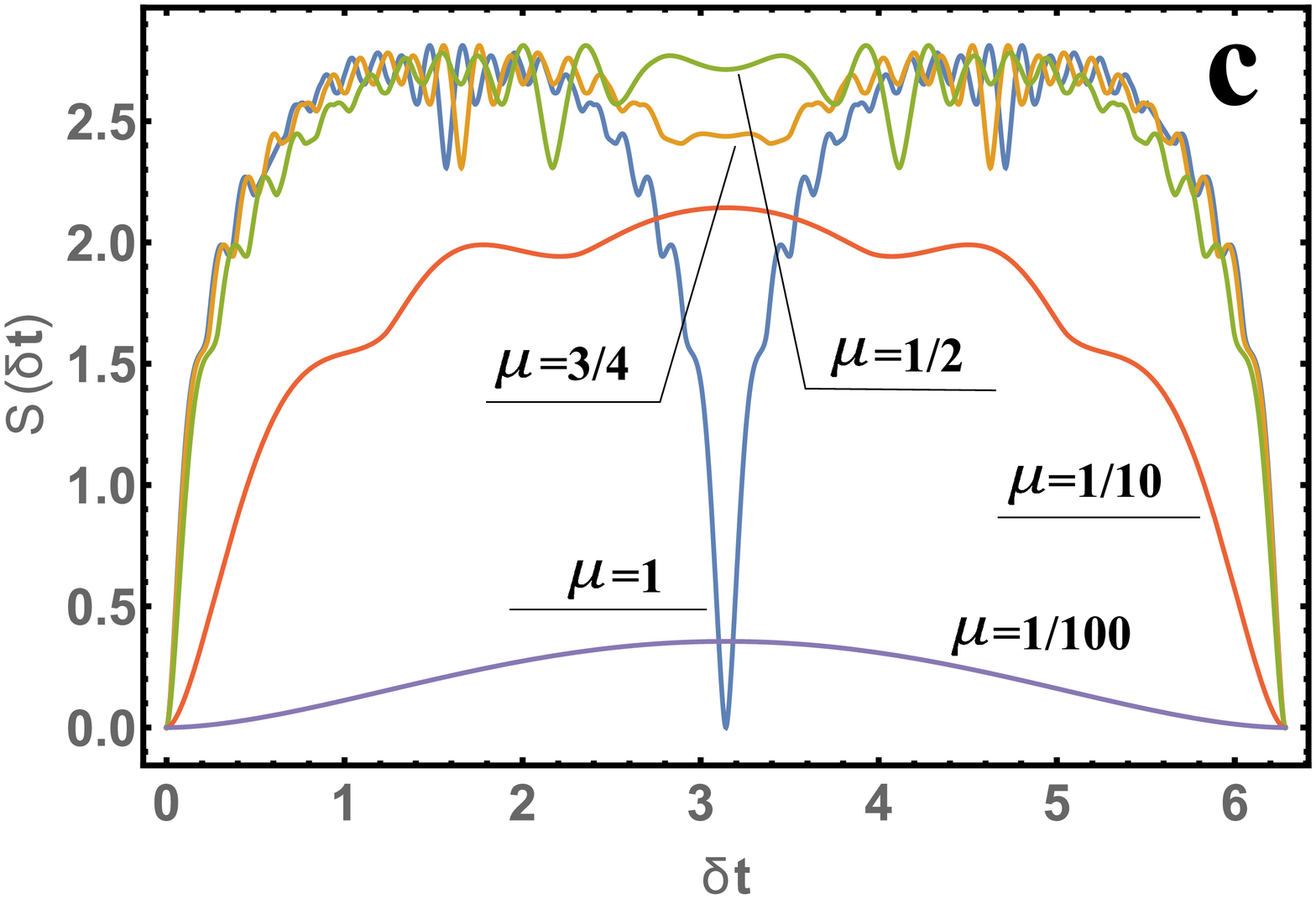}} \\
\end{minipage}
\hfill
\begin{minipage}[h]{0.49\linewidth}
\center{\includegraphics[angle=0,width=1\textwidth, keepaspectratio]{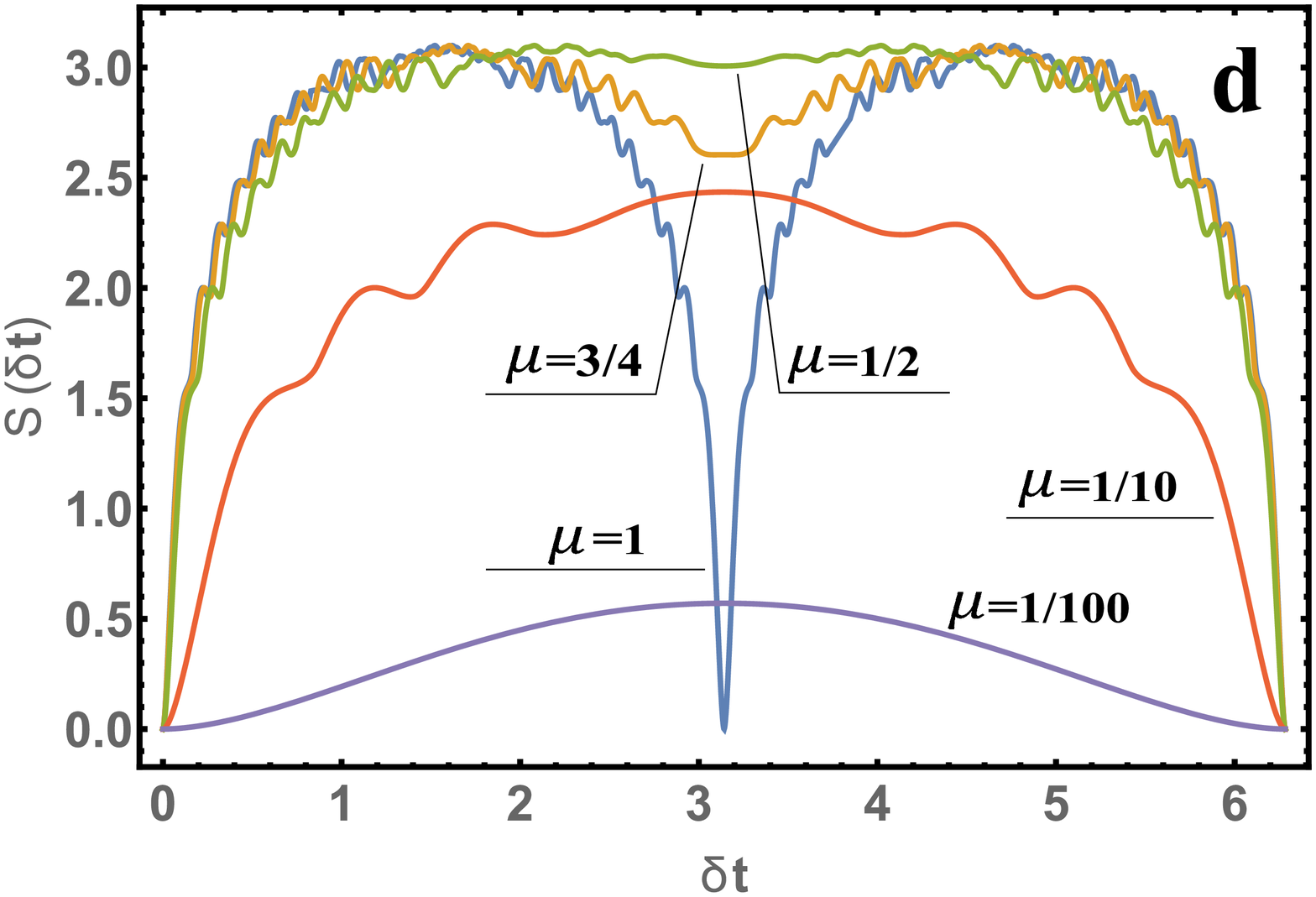}} \\
\end{minipage}
\center{\caption[fig]{ The results of calculating the Von Neumann entropy $S=S(\delta t)$ for  $\mu = (1,3/4,1/2,1/10,1/100)$ and a) $s_1=0,s_2=10$; b) $s_1=5,s_2=10$; c) $s_1=10,s_2=10$; d) $s_1=20,s_2=10$  }}
\label{fig}
\end{figure}
It should be added that when calculating the quantum entanglement using the Schmidt mode $\lambda_k(t)$, the expression for $\Delta E_{n,s_{1}+s_{2}-n} $ (see (\ref{18} )) can be replaced by $\Delta E_{n,s_{1}+s_{2}-n} \to \delta n $, where $ \delta = \frac {C(\epsilon + \mu)} { \sqrt{A^{'}} + \sqrt{B^{'}}} $. It can also be observed that $ \lambda_k(t) $ is a smooth function, in spite of the fact that $\epsilon=0 $ has a discontinuity at $\epsilon=0$. We represent the results of the quantum entanglement of the system under consideration using the Von Neumann entropy $ S $. As an example, in Figure 2 we consider the dependence of $ S=S(\delta t) $ on the dimensionless parameter $ \delta t $, with $ \mu = (1,3/4,1/2,1/10,1/100) $ and four variants $ s_1, s_2 $.

\section{Conclusion}

A system consisting of two coupled harmonic oscillators has been studied. This work has demonstrated that it is possible to obtain in an analytical form the main characteristics of the quantum entanglement of a system, for example, the Von Neumann entropy or the Schmidt parameter. The Hamiltonian considered is of great interest for various branches of physics, chemistry, and biology particularly because the obtained analytical expressions can be used for analysis and corresponding conclusions without resorting to numerical calculations that are difficult to undertake for large values of quantum numbers. In addition, it has been shown that quantum entanglement can be very large, which is a result that  can be used in various fields of quantum informatics and other fields in physics, for example, in models that are examples of the remainder of the Feynman universe.

\end{document}